\documentclass[%
reprint,
nofootinbib,
amsmath,amssymb,
prl,
]{revtex4-1}

\usepackage[export]{adjustbox}
\usepackage{color}
\usepackage{tikz}
\usepackage{graphicx}% Include figure files
\usepackage{dcolumn}% Align table columns on decimal point
\usepackage{bm}% bold math
\usepackage{hyperref}% add hypertext capabilities
\usepackage{float}
\usepackage[caption=false]{subfig}
\usepackage{hyperref}
\hypersetup{
    colorlinks=true,
    linktoc=page,    %set to all if you want both sections and subsections linked
    linkcolor=blue,
    urlcolor=magenta
}

\newcommand{\be}{\begin{equation}}
\newcommand{\ee}{\end{equation}}
\newcommand{\bea}{\begin{eqnarray}}
\newcommand{\eea}{\end{eqnarray}}
\newcommand{\dd}{\text{d}}
\newcommand{\bra}[1]{\left<{#1}\right|}
\newcommand{\ket}[1]{\left|{#1}\right>}
\newcommand{\ave}[1]{\left<{#1}\right>}
\newcommand{\tN}{\text{N}}
\newcommand{\tH}{\text{H}}
\newcommand{\hh}{\text{h}}
\newcommand{\hH}{\text{H}}
\newcommand{\tE}{\text{E}}
\newcommand{\tr}{\text{tr}}

\begin{document}

\title{Black hole thermodynamics from an ensemble-averaged theory}

\author{Peng Cheng$^{1}$}
\email{pchengcn@tju.edu.cn}
\author{Yu-Xiao Liu$^{2}$}
\email{liuyx@lzu.edu.cn}
\author{Shao-Wen Wei$^{2}$}
\email{weishw@lzu.edu.cn}

\affiliation{1) Center for Joint Quantum Studies and Department of Physics, School of Science,
Tianjin University, Tianjin 300350, China\\
2) Lanzhou Center for Theoretical Physics, Key Laboratory of Theoretical Physics of Gansu Province, and Key Laboratory of Quantum Theory and Applications of MoE, Lanzhou University, Lanzhou, Gansu 730000, China}

\begin{abstract}
The path integral approach to a quantum theory of gravity 
is widely regarded as an indispensable strategy.
However, determining what additional elements, beyond black hole or AdS spacetime, should be incorporated into the path integral remains crucial yet perplexing.
We argue that the spacetime with a conical singularity in its Euclidean counterpart should be the most important ingredient to append to the path integral.
Therefore, physical quantities should be ensemble-averaged over all geometries since they are described by the same Lorentzian metric.
When the ensemble average is introduced, the Hawking-Page transition for the Schwarzschild-AdS black hole and the small-large black hole transition for the Reissner-Nordstr\"{o}m-AdS black hole naturally arise as semi-classical approximations, when the size of the black hole system is much larger than the Planck length.
Away from the semi-classical limit, the system is a superposition of different geometries, and the averaged quantities would deviate from the black hole thermodynamics.
Expanding around the classical saddles, the subleading order of the Newton constant contributions can be derived, which are half of the Hawking temperature both for the Schwarzschild and Reissner-Nordstr\"{o}m black holes. The result may imply a universal structure.
The subsubleading terms and more intriguing physics that diverge from black hole thermodynamics are revealed.
The ensemble-averaged theory provides a new way of studying subleading effects and extending the traditional AdS/CFT correspondence.
\end{abstract}

\maketitle
\flushbottom

%%%%%%%%%%%%%%%%%%%%%%%%%%%%%%%%%%%%%%%%%%%%%%%%%%%%%%%%%%%%%%%%%%%%%%%%%%%%%%%%%%%%%%%%%%%%%%%%%%%%
% MAIN BODY
%%%%%%%%%%%%%%%%%%%%%%%%%%%%%%%%%%%%%%%%%%%%%%%%%%%%%%%%%%%%%%%%%%%%%%%%%%%%%%%%%%%%%%%%%%%%%%%%%%%%

%\tableofcontents

\section{Introduction}
\label{intro}

Black hole phase transition has always been an intriguing subject, providing us with many profound insights.
As a prime example of the AdS/CFT correspondence, the Hawking-Page phase transition demonstrating a transition between thermal radiation in AdS spacetime and black holes \cite{Hawking:1982dh}, has a boundary duality which is a large-$N$ SU($N$) gauge theory on a sphere.
It was interpreted as a phase transition between a confined phase with energy $E\propto N^0$ and a deconfined phase with energy $E\propto N^2$ in the dual field theory \cite{Witten:1998zw,Gross:1980he,Sundborg:1999ue}.
For charged AdS black holes, it was demonstrated that there is a small-large black hole phase transition \cite{Chamblin:1999hg,Chamblin:1999tk,Kastor:2009wy,Dolan:2011xt,Kubiznak:2012wp}.
These phase transitions provide a window for understanding the microstates of black holes \cite{Wei:2015iwa}. Furthermore, the free energy landscape proposal was developed to understand different phase transitions based on the generalized free energy \cite{Li:2020khm,Li:2020nsy,Li:2020spm,Yang:2021ljn,Li:2021vdp,Li:2022oup,Li:2022yti,Li:2022ylz}.

The key assumption of black hole thermodynamics is that black holes are thermodynamical systems, and the states with lower free energies are preferred, which is the origin of phase transitions.
However, it is hard to associate any microstructure to a null horizon in general relativity, and a phase transition between small and large black holes is rather extraordinary to imagine.
In black hole thermodynamics, 
we usually start by assuming that we deal with a thermodynamic system, and then examine how similarities can be drawn between black holes and such systems. However, this reasoning is evidently circular.
Then, it is natural to ask whether there exists a more fundamental method for deriving the phase transition and to determine the conditions under which black hole thermodynamics holds true. Moreover, a statistical interpretation of the black hole thermodynamics is crucial but remains unclear.

\begin{figure}[hbt]
  \includegraphics[width=5.5cm]{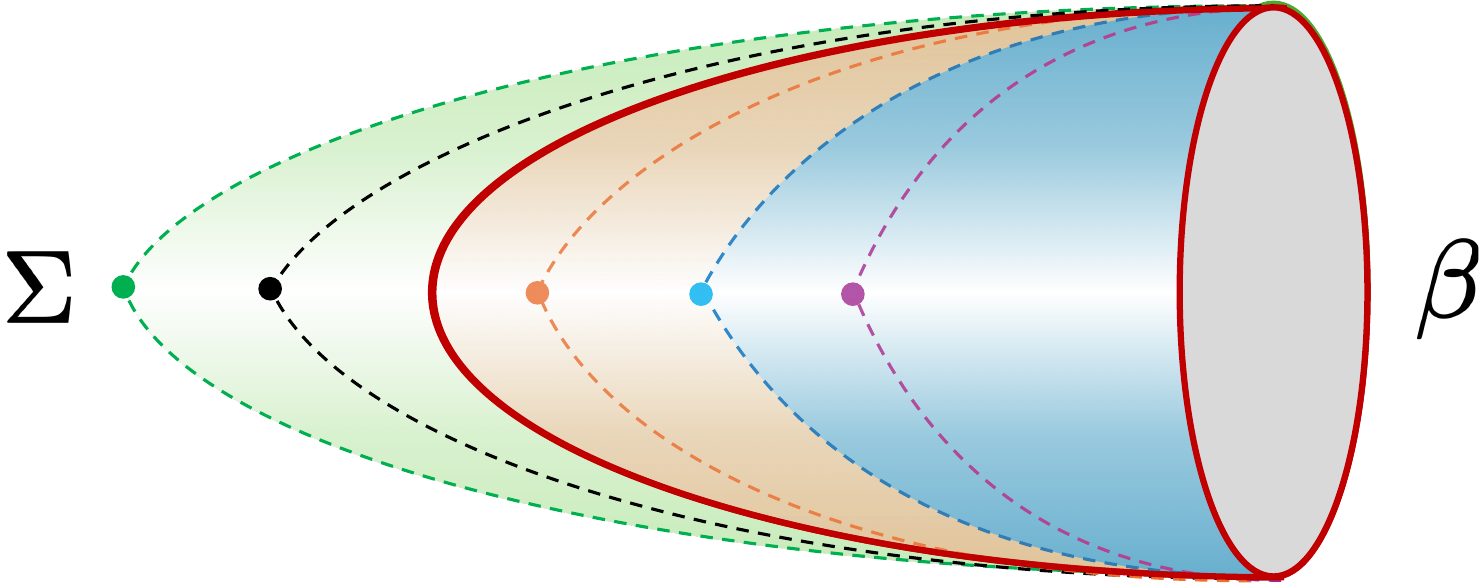}
  \caption{Schematic of the ensemble average for Euclidean Schwarzschild geometries. The orange ``cigar'' geometry represents the Euclidean Schwarzschild black hole, while the other geometries represented by dashed curves possess conical singularities at their tips. We propose that all the physical quantities should be ensemble-averaged over those geometries, weighted by the corresponding gravitational action.
   }\label{Euc}\vspace{-1em}
\end{figure}

To address the above problem, we will introduce a method called ensemble average. In this method, all physical quantities are ensemble averaged over different geometries, as displayed in Fig. \ref{Euc} for Euclidean Schwarzschild geometries. 
Let us assume the corresponding Hawking temperature of the Schwarzschild black hole is $T_\tH=1/\beta_\tH$. When the periodicity of Euclidean time deviates from $\beta_\tH$, a conical singularity will occur. 
In the Euclidean path integral for quantum gravity, all possible geometries including those with $\beta \neq \beta_\tH$, should be included in and weighted by the action.  
Moreover, physical quantities related to the Schwarzschild metric should be an ensemble average over all the geometries, because all the geometries shown in Fig. \ref{Euc} are described by the same Lorentzian metric.
When an ensemble average over all the above geometries is introduced, we would see that many intriguing properties naturally arise. 
Note that there are further motivations for introducing ensemble average from low-dimensional gravity, where the concept was introduced to understand the factorization puzzle \cite{Saad2019,AfkhamiJeddi2020,Maloney2020,Penington2019a,Marolf2020a,Maldacena2016a,Jensen2016,Heckman:2021vzx,Cheng:2022nra,Jafferis:2024jkb}.

The idea of including off-shell geometries with conical singularity in Euclidean path integral is a widely studied one. 
Following Gibbons and Hawking's original proposal of Euclidean quantum gravity \cite{Gibbons:1976ue}, 
York and his collaborators conducted a series of studies examining the situation when temperature variations are permitted \cite{York:1986it,Braden:1987ad,Whiting:1988qr,York:1988va}.
Especially, the so-called generalized free energy and related Euclidean action were first studied in \cite{York:1986it}.
However, in those studies, the black hole was argued to be placed in a cavity, and the asymptotic AdS spacetime was not considered.
Subsequently, Solodukhin and his collaborators treated the geometries with conical singularity more seriously \cite{Solodukhin:1994yz,Fursaev:1995ef,Mann:1996bi,Solodukhin:2011gn}. They also considered correction to the classical result. However, their correction is from the scalar field living on the fixed geometry rather than the correction due to the conical geometry itself in the Euclidean path integral.
In Ref. \cite{Carlip:1993sa}, Carlip and Teitelboim emphasized the importance of the off-shell black holes due to the conical singularity and pointed out that understanding the weight of each state is crucial to the subject. 
The current letter exactly deals with the weight of each geometry in the ensemble-averaged theory.
The renascence of the topic is mainly because of the free energy landscape proposal 
\cite{Li:2020khm,Li:2020nsy,Li:2020spm,Yang:2021ljn,Li:2021vdp,Li:2022oup,Li:2022yti,Li:2022ylz}, where the kinetics of the black hole phase transition was studied in terms of the generalized free energy.

%It was proposed that wormholes can be understood as the variance of the averaging, such that factorization is not a necessary property both in the bulk and boundary. 
%If one is looking for a one-to-one correspondence between the bulk and boundary, the bulk gravity should also be regarded as an ensemble average. 

In this letter, we aim to understand black hole physics by introducing the concept of the ensemble average over all geometries on the free energy landscape, and by studying the effects deviating from the classical saddles in the gravitational path integral. 
The semi-classical limit of the ensemble-averaged physics is just the black hole thermodynamics, where we have the Hawking-Page and small-large black hole phase transitions.
The new frame allows us to analyze the quantum gravitational corrections to black hole physics. 
We study the subleading orders of $G_\tN$ expansions of the Schwarzschild-AdS and Reissner-Nordstr\"{o}m (RN)-AdS black hole, from which we can see the universal structures of the quantum corrections.

%\vspace{-5em}
\section{Black hole thermodynamics and ensemble average}\label{averaging}

The Schwarzschild black hole in asymptotic AdS spacetime is regarded as a thermodynamical system. 
%The corresponding metric is
%\be
%\dd s^2=-f(r)\dd t^2+f(r)^{-1}\dd r^2+r^2(\dd \theta^2+\sin^2\theta \dd \phi^2)\label{Smetric}
%\ee
%where
%\be
%f(r)=1-\frac{2G_\tN M}{r}+\frac{r^2}{L^2}\,.\nonumber
%\ee
The temperature $T_\tH$, entropy $S$ and free energy $F$ of the system can be expressed as \cite{Bardeen:1973gs,Bekenstein:1973ur,Hawking:1975vcx,Hawking:1976de}
\bea
T_\tH &=& \frac{1}{4 \pi  r_\hH}\left(1+\frac{3 r_\hH^2}{ L^2}\right)\,,~~~~S = \frac{\pi  r_\hH^2}{G_\tN}\,,\\
F &=& M-T_\tH S=\frac{r_\hH}{4G_\tN}\left(1-\frac{r_\hH^2}{L^2}\right)\,.\label{BHfree}
\eea
$M$ is the mass of the black hole, $r_{\hH}$ is the black hole horizon radius, and $L$ is the AdS radius.
%\bea
%T_\tH = \frac{1}{4 \pi  r_\hh}\left(1+\frac{3 r_\hh^2}{ L^2}\right)\,~~S = \frac{\pi  r_\hh^2}{G_\tN}\,~~F =\frac{r_\hh}{4G_\tN}\left(1-\frac{r_\hh^2}{L^2}\right)\,.\nonumber\\
%\label{BHfree}
%\eea
When the system is regarded as thermodynamical, a state with lower free energy is thermodynamically preferred. 
So, there is a phase transition between the thermal AdS and black holes at Hawking-Page temperature $T_{\text{HP}}$.
The transition is called the Hawking-Page transition. 
In black hole thermodynamics, varying $G_\tN$ does not influence the phase diagram, because only saddle points are considered, and $G_\tN$ has no contributing effect.

In the path integral approach to a quantum theory of gravity, one should integrate over all possible geometries weighted by action $S[g]$. 
The partition function can be evaluated via Euclidean path integral, with Euclidean geometries and Euclidean action $I_\tE[g]$ \cite{Gibbons:1976ue}. 
Since all the geometries shown in Fig. \ref{Euc} are described by the same Lorentz metric, which is a solution of Einstein's equation, they should be included in the path integral. The question now lies on what are their corresponding actions.

Euclidean geometries with conical singularities at their tips were discussed in \cite{Fursaev:1995ef,Solodukhin:1994yz,Mann:1996bi,Solodukhin:2011gn,Li:2022oup}. 
For the Euclidean Schwarzschild-AdS case, the Euclidean Einstein-Hilbert action is given by
\be
I_\tE=-\frac{1}{16\pi G_\tN}\int_{\mathcal{M}}\left(R+\frac{6}{L^2}\right)\sqrt{g}~\dd^4x\,.
\ee
When considering a geometry with a conical singularity $\Sigma$ as shown in Fig. \ref{Euc}, the action contains an extra contribution from the conical singularity
\be
I_{\Sigma}%=-\frac{1}{4G_\tN} \left(1-\frac{\beta}{\beta_\tH}\right)\int_{\mathcal{H}} 1
=-\left(1-\frac{\beta}{\beta_\hh}\right)\frac{\pi r_\hh^2}{G_\tN}\,.\label{IS}
\ee
$r_\hh$ is the horizon radius corresponding to the horizon temperature $1/\beta_{\hh}$. After the regularization procedure, the final Euclidean action can be evaluated as \cite{Li:2022oup}
\be
I_\tE[\text{S-AdS}]=\frac{\beta r_\hh}{2G_\tN}\left(1+\frac{r_\hh^2}{L^2}-2\pi  r_\hh T\right)\,.\label{ISAdS}
\ee
Note that as can be seen from \eqref{IS}, when the ensemble temperature equals the horizon temperature, i.e. $\beta=\beta_\hh$, the extra term vanishes, and we get the Euclidean action of the black hole. 
So the actions for geometries without conical singularity are also calculated.

\begin{figure*}[hbt]
\centering
  \subfloat[Averaged free energy for the Schwarzschild-AdS spacetime. \label{SAdS}]{\includegraphics[width=7.3cm]{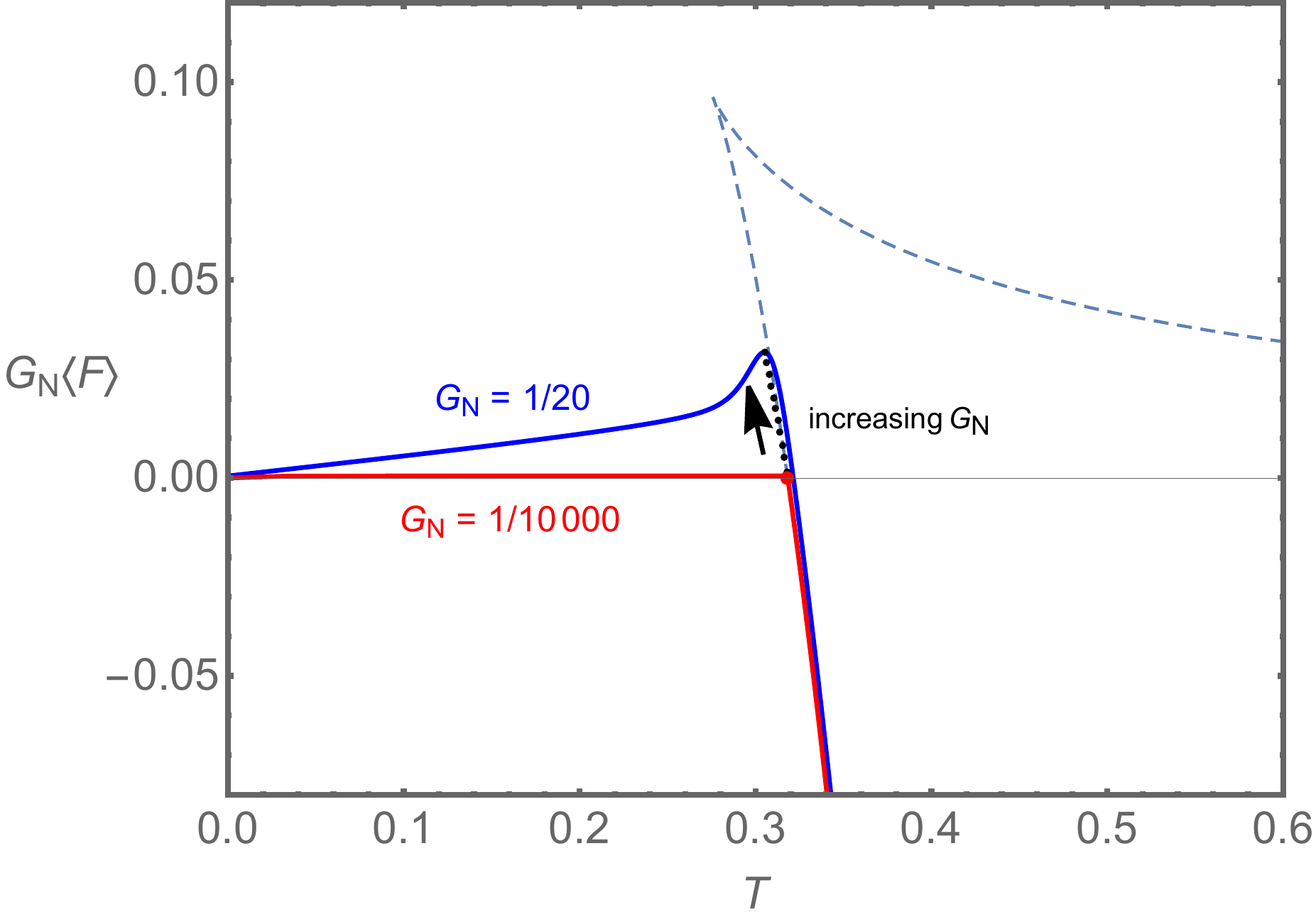}}
~~~~~~~\subfloat[Averaged free energy for the RN-AdS spacetime. \label{RN-AdS}]{\includegraphics[width=7cm]{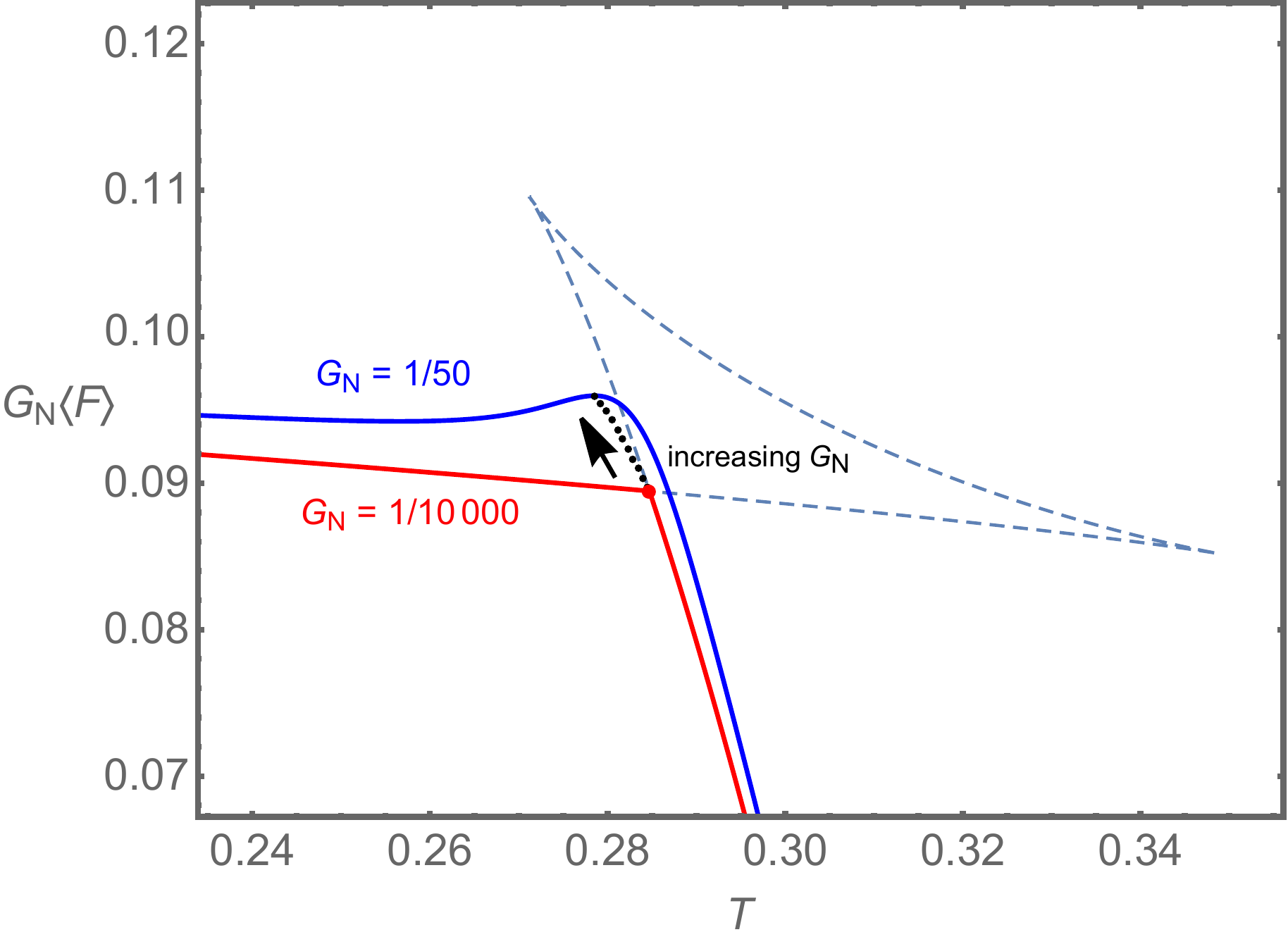}}
  \caption{
 Averaged free energy $\ave{F}$ for the Schwarzschild-AdS and RN-AdS spacetimes.
 (a) Averaged free energy for the Schwarzschild-AdS spacetime. The blue and red curves represent $\ave{F}$ with $G_\tN=1/20$ and $G_\tN=1/10000$.  
 (b) Averaged free energy for the RN-AdS spacetime. The blue and red curves represent the situation with $G_\tN=1/50$ and $G_\tN=1/10000$. 
 The Hawking-Page transition and small-large black hole transitions can be regarded as a small $G_\tN$ approximation of the ensemble-averaged physics, illustrated by the red curves.
 While for relatively large $G_\tN$, we do not have a sharp phase transition. The $G_\tN$ dependence of transition points is indicated by the arrows.\\
  } \label{ave}\vspace{-2em}
\end{figure*}

With the Euclidean action, we must include all those geometries with weight $e^{-I_\tE}$ in the path integral. 
Moreover, for any physical quantity associated with a certain Lorentzain metric, we should define it as an ensemble-averaged quantity over all the possible Euclidean geometries shown in Fig. \ref{Euc}.
The basic principle for defining the ensemble average is that we fix boundary ensemble temperature $T=1/\beta$ and average over the bulk geometries with the same Lorentzian metric but with different horizon temperatures. 
The ensemble average must be based on the canonical coordinate and conjugate momentum. 
However, the Euclidean action in \eqref{ISAdS} is a potential, and we need to consider the perturbative kinetic energy on the potential to determine what variables to integrate in the ensemble average. 
From \cite{Li:2021vdp,Li:2022yti,Li:2022ylz}, the conjugate momentum is $\dot{r}_\hh$ and it is natural to use $r_\hh$ as the canonical coordinate.
Integrating over $\dot{r}_\hh$ results in a constant, which is canceled by the same factor from the denominator. 
So, we can use $r_\hh$ to characterize different geometries, and define ensemble-averaged quantity $\ave{A}$ as
%\be
%\ave{A}=\frac{\int A(r_\hh) ~e^{-I_\tE(r_\hh)}\dd r_\hh}{\int e^{-I_\tE(r_\hh)}\dd r_\hh}\,.
%\ee
\be
\ave{A}=\frac{1}{Z}~\tr(\rho A)\,,\label{aveA}
\ee
For example, the free energy corresponding to the Schwarzschild metric should be an ensemble-averaged quantity
%\be
%\ave{F}%=\frac{\int \frac{I_\tE}{\beta} ~e^{-I_\tE}\dd r_\hh}{\int e^{-I_\tE}\dd r_\hh}
%=\frac{\int \frac{ r_\hh}{2G_\tN}\left(1+\frac{r_\hh^2}{L^2}-2\pi  r_\hh T\right) ~e^{-I_\tE}\dd r_\hh}{\int e^{-I_\tE}\dd r_\hh}\,.
%\ee
\be
\ave{F}= \frac{1}{Z}~\tr(\rho F_{\text{gen}}) = \frac{\int \frac{ r_\hh}{2G_\tN}\left(1+\frac{r_\hh^2}{L^2}-2\pi  r_\hh T\right) ~e^{-I_\tE}\dd r_\hh}{\int e^{-I_\tE}\dd r_\hh}\,.\label{aveF}
\ee
The ensemble average inherits the spirit of including all the possible geometries in the path integral. $e^{-I_\tE}$ can be regarded as the probability of each geometry, and $\int e^{-I_\tE}\dd r_\hh$ in the denominator of \eqref{aveF} is the normalization factor. %see appendix \ref{appA} for more details.

The Euclidean action for the RN-AdS case can be derived through a similar method \cite{Li:2022oup}
\be
I_\tE[\text{RN-AdS}]=\frac{\beta r_\hh}{2G_\tN} \left(1+\frac{r_\hh^2}{L^2}+\frac{Q^2}{r_\hh^2}-2\pi  r_\hh T\right)\,
\ee
with the event horizon radius $r_\hh$. All the physical quantities related to the RN-AdS metric should also be defined as ensemble-averaged ones over all the Euclidean geometries with or without conical singularities.

\section{Black hole phase transition as a small $G_\tN$ effect}
\label{phase}

\begin{figure*}[hbt]
\centering
\subfloat[The Schwarzschild-AdS case.
\label{subS}]{\includegraphics[width=7cm]{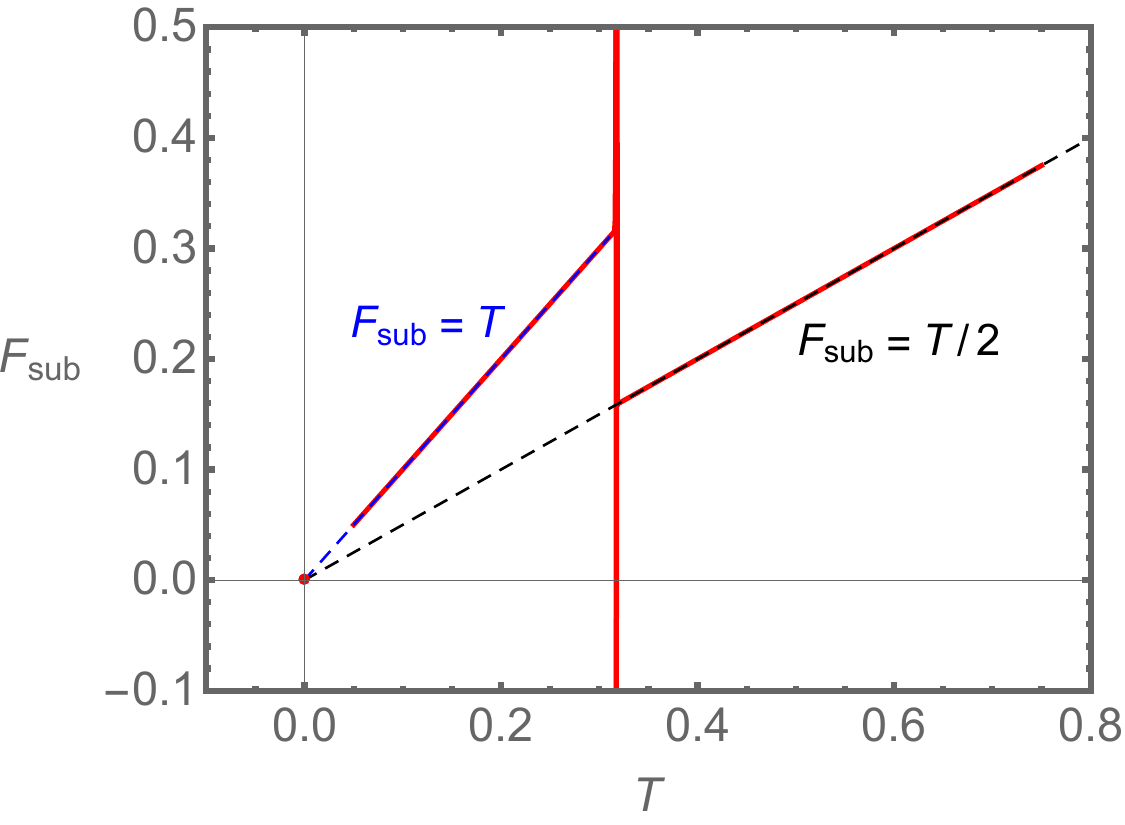}}
~~~~~~~~\subfloat[The RN-AdS case.
\label{subRN}]{\includegraphics[width=7cm]{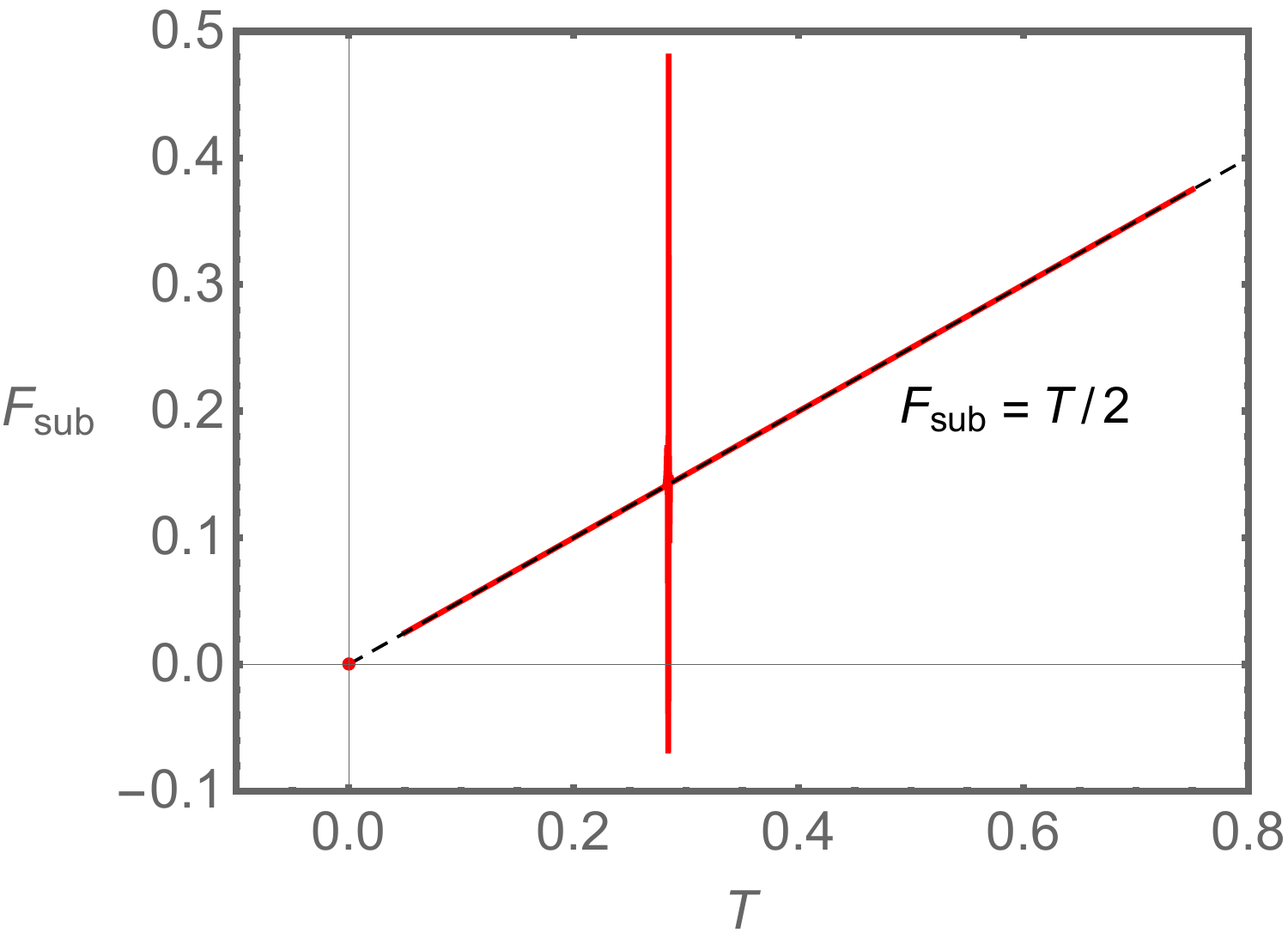}}
\caption{Subleading-order contributions of the Schwarzschild-AdS and RN-AdS spacetimes. (a) The subleading contribution of the thermal AdS matches $F_{\text{sub}}=T$ shown by the blue dashed line. While the subleading contribution of the Schwarzschild-AdS is $F_{\text{sub}}=T/2$. (b) Before or after the phase transition, the subleading contribution of the RN-AdS is always $F_{\text{sub}}=T/2$.}
\label{sub}\vspace{-1em}
\end{figure*}

The Euclidean action for the Schwarzschild-AdS black hole can be rewritten as
\be
I_\tE[\text{S-AdS}]
=\frac{\beta r_\hh}{2 l_{\text{pl}}^2}\left(1+\frac{r_\hh^2}{L^2}-2\pi  r_\hh T\right)\,.\label{SS}
\ee
Inspired by the holographic dictionary where the bulk semi-classical limit is defined as small $l_{\text{pl}}^{d-2}/L^{d-2}$ limit, we can use the AdS radius $L$ as the unit length scale and define the dimensionless Newton constant $\tilde{G}_\tN$ as
\be
\tilde{G}_\tN=l_{\text{pl}}^2/L^2\,.
\ee
Then, the Euclidean action \eqref{SS} can be written as
\be
I_\tE[\text{S-AdS}]
=\frac{1}{2\tilde{G}_\tN}\frac{\beta r_\hh}{L^2}\left(1+\frac{r_\hh^2}{L^2}-2\pi  r_\hh T\right)\,.\label{tG}
\ee
%we can use the AdS radius $L$ as the unit length scale, and rewrite all the scales in the system as dimensionless parameters, for example
%\bea
%\tilde{G}_\tN=l_{\text{pl}}^2/L^2\,,~~\tilde{\beta}=\beta/L\,,~~\tilde{r}_{\hh}=r_{\hh}/L\,,~~\tilde{L}=1\,.\nonumber
%\eea
%Then, the Euclidean action \eqref{SS} can be written as 
%\be
%I_\tE[\text{S-AdS}]
%=\frac{\tilde{\beta} \tilde{r}_\hh}{2 \tilde{G}_\tN}\left(1+\frac{\tilde{r}_\hh^2}{\tilde{L}^2}-2\pi  \tilde{r}_\hh \tilde{T}\right)\,.
%\ee
The so-called semi-classical regime is the situation when $\tilde{G}_\tN\ll 1$, which also means $l_{\text{pl}}^2\ll L^2$. For large $\tilde{G}_\tN$, quantum corrections should be included. From now on, we will drop the tilde sign and regard $G_\tN\ll 1$ as the semi-classical limit.
Whenever considering quantum gravitational corrections, we should be able to recover the black hole thermodynamics in the semi-classical limit. In this section, we will analyze the semi-classical limit of the ensemble-averaged theory and study the deviation from the thermodynamics due to the effects of non-saddle geometries.

The averaged free energies $\ave{F}$ for the Schwarzschild-AdS and RN-AdS spacetimes with varying $G_\tN$ values are plotted against the ensemble temperature $T$ in Fig. \ref{SAdS} and Fig. \ref{RN-AdS}, respectively. 
As indicated by the red curves, the Hawking-Page transition between thermal AdS and a large black hole for Schwarzschild-AdS, and the small-large black hole transition for RN-AdS naturally arise as a result of the small $G_\tN$ limit. 
In black hole thermodynamics, the phase transition is a switch of dominant saddles in the Euclidean path integral. The small $G_\tN$ limit implies the effects away from the saddles can be ignored and we only have a summation of isolated saddles.

Note that for small $G_\tN$, the thermal AdS geometry naturally arises. This is much less artificial than the Schwarzschild black hole thermodynamics reviewed at the beginning of the previous section, where one needs to separately evaluate and compare the free energy for the thermal AdS and black hole. 
While the phase transitions naturally arise in the small $G_\tN$ limit in our case.

While for relatively large $G_\tN$, there can be non-ignorable contributions from the Euclidean geometries with conical singularities. 
The non-black-hole geometries included according to the probability $e^{-I_\tE}$ would have significant influences on black hole physics, and the effect is demonstrated by the deviation of the averaged free energy from the black hole free energy.
As can be seen from the blue curves in Fig. \ref{ave}, the averaged free energy does not exactly match the black hole free energy, and we do not have sharp phase transitions.
This phenomenon makes sense in the context of holography duality because finite $1/G_\tN$ corresponds to finite $N$ in the boundary theory. There would not be a sharp phase transition for a system with finite degrees of freedom.
The point with maximal averaged free energy can be more or less regarded as a ``transition point'' for different phases, because the free energy, after initially increasing, begins to decrease after the point.
The transition points for different $G_\tN$ are illustrated by black dots in Fig. \ref{ave} both for the Schwarzschild-AdS and RN-AdS cases. 
With increasing $G_\tN$ the trends of the deviations are indicated by black arrows in the figures.

We have some comments on the relation between the ensemble-averaged theory and the black hole free energy landscape. In the free energy landscape proposal \cite{Li:2020khm,Li:2020nsy,Li:2020spm,Yang:2021ljn}, the topography of the generalized free energy is used to understand the black hole phase transition. Local minimums of the generalized free energy have probabilities of switching to a global one. When this happens, there would be a phase transition. Moreover, the process is understood as stochastic motions in the topographic map and described by the stochastic Fokker-Planck equation.
The ensemble-averaged theory proposed here is inspired by the free energy landscape, and we suggest all the geometry (the whole topographic map) should be integrated over with the probability inherited from the gravitational path integral.
This provides a natural explanation of the topography and probability in terms of the Euclidean path integral.
The free energy landscape is also helpful in understanding our results. For small $G_\tN$, the valley of the landscape is super deep. The deepness of the global minimum is presented by the crest of the probability $e^{-I_\tE}$. Thus, we can only see the contribution from the global minimum.
As will be seen in the next section, to see the quantum corrections to the black hole thermodynamics one can expand the Euclidean action and the averaged free energy around classical saddles.

\section{Non-classical corrections to black hole thermodynamics}
\label{quantum}

To further understand the quantum effects due to the new geometries, we can do the semi-classical expansion of the averaged free energy \eqref{aveF}.
In the Supplemental Material, it is shown that for small $G_\tN$, the probability distribution of each state can be approximated by a Gaussian distribution in black hole phases. For more discussions, see \cite{Cheng:2024efw,Ali:2024adt}. 
%So we can expand around the classical saddle to the second order and use a Gaussian distribution to study the ensemble-averaged theory.
%When the horizon temperature equals the ensemble temperature, we have a bulk black hole without conical singularity whose horizon radius is denoted as $r_\hH$.
So we can expand $I_\tE$ up to the second order of $\Delta=r_\hh-r_\hH$ near $r_\hh=r_\hH$, and integrate over the neighborhood of $r_\hH$ in \eqref{aveF}. 
Note that $r_\hH$ means the horizon radius of the ``Hawking saddle'' \cite{Penington2019a}, while $r_\hh$ is a free parameter.
Expanding $I_\tE$ naturally gives us a Gaussian distribution of different geometries. The second-order term corresponds to the $\mathcal{O}(G_\tN^0)$ contribution in the averaged free energy.
For small $G_\tN$, the probability distribution $e^{-I_\tE}$ is very steep near the saddle point, and the Gaussian integral can run from $(r_\hH-5\sigma,r_\hH+5\sigma)$ with variance $\sigma^2\propto G_\tN$. 
The error function of the Gaussian integral can be evaluated, which would be very small.

Expanding the Euclidean action and the free energy to the second order, we have
\bea
I_\tE^{\Delta} &=& \frac{\beta r_\hH}{2G_\tN}\left(1+\frac{r_\hH^2}{L^2}-2\pi  r_\hH T\right)+\frac{3\beta r_\hH-2\pi L^2}{2G_\tN L^2}\Delta^2\,,\nonumber\\
F^{\Delta} &=& \frac{r_\hH}{2G_\tN}\left(1+\frac{r_\hH^2}{L^2}-2\pi  r_\hH T\right)+\frac{3r_\hH-2\pi L^2 T}{2G_\tN L^2}\Delta^2\,.\nonumber
\eea
The coefficient of the $\Delta^2$ term can be identified as $1/(2\sigma^2)$. The averaged free energy can be evaluated as
\bea
\ave{F}&\approx&\frac{\int_{-5\sigma}^{+5\sigma} F^{\Delta}e^{-I_\tE^{\Delta}}\dd \Delta }{\int_{-5\sigma}^{+5\sigma} e^{-I_\tE^{\Delta}}\dd \Delta }\nonumber\\
&=& \frac{r_\hH}{4G_\tN}\left(1-\frac{r_\hH^2}{L^2}\right)+\frac{T}{2}+\frac{5 e^{-\frac{25}{2}}}{\sqrt{2 \pi }  ~\text{erf}\left(\frac{5}{\sqrt{2}}\right)}T.\label{GaussBH}
\eea
As expected, the leading-order contribution is just the black hole free energy shown in \eqref{BHfree}. 
The subleading contribution of order $\mathcal{O}(G_\tN^0)$ can be approximated as $F_{\text{sub}}^{\text{BH}}=T/2$.
The last error term due to integrating over $(-5\sigma,5\sigma)$ in \eqref{GaussBH} is $7.4\times 10^{-6}\times T$, thus can be ignored.
%The error term is even smaller when integrating over a region larger than $(-5\sigma,5\sigma)$.
The subleading-order contribution for the RN-AdS case can be derived by a similar method, and we also have $F_{\text{sub}}^{\text{BH}}=T/2$.
Note that we expand the free energy near the saddle point, where the ensemble temperature $T$ should be approximate to the Hawking temperature $T_\tH$ due to small $\sigma\propto G_\tN$ for small $G_\tN$. So we have a corrected free energy with subleading correction roughly equals $T_\tH/2$. 

Let us consider the thermal AdS case which can be regarded as the zero $r_\hh$ limit of the Schwarzschild-AdS black hole. With a deformation $\Delta$, the Euclidean action and free energy can be denoted as
\be
I_\tE^{\Delta}=\frac{\beta\Delta}{2G_\tN}\,,~~~~F^{\Delta}=\frac{\Delta}{2G_\tN}\,.
\ee
The averaged free energy can be calculated near the AdS saddle as
\be
\ave{F}\approx\frac{\int_{0}^{5 \sqrt{G_\tN}} F^{\Delta} ~e^{-I^{\Delta}_\tE}\dd \Delta }
{\int_{0}^{5 \sqrt{G_\tN}} e^{-I^{\Delta}_\tE}\dd \Delta } 
=T\,.
\ee
The error term can always be ignored because $\beta\gg \sqrt{G_\tN}$ always guarantees the error is super small.
So for the thermal AdS case, the $1/G_\tN$-order contribution is zero, and the subleading contribution can be derived as $F_{\text{sub}}^{\text{AdS}}=T$.

\begin{figure}[hbt]
\includegraphics[width=6.5cm]{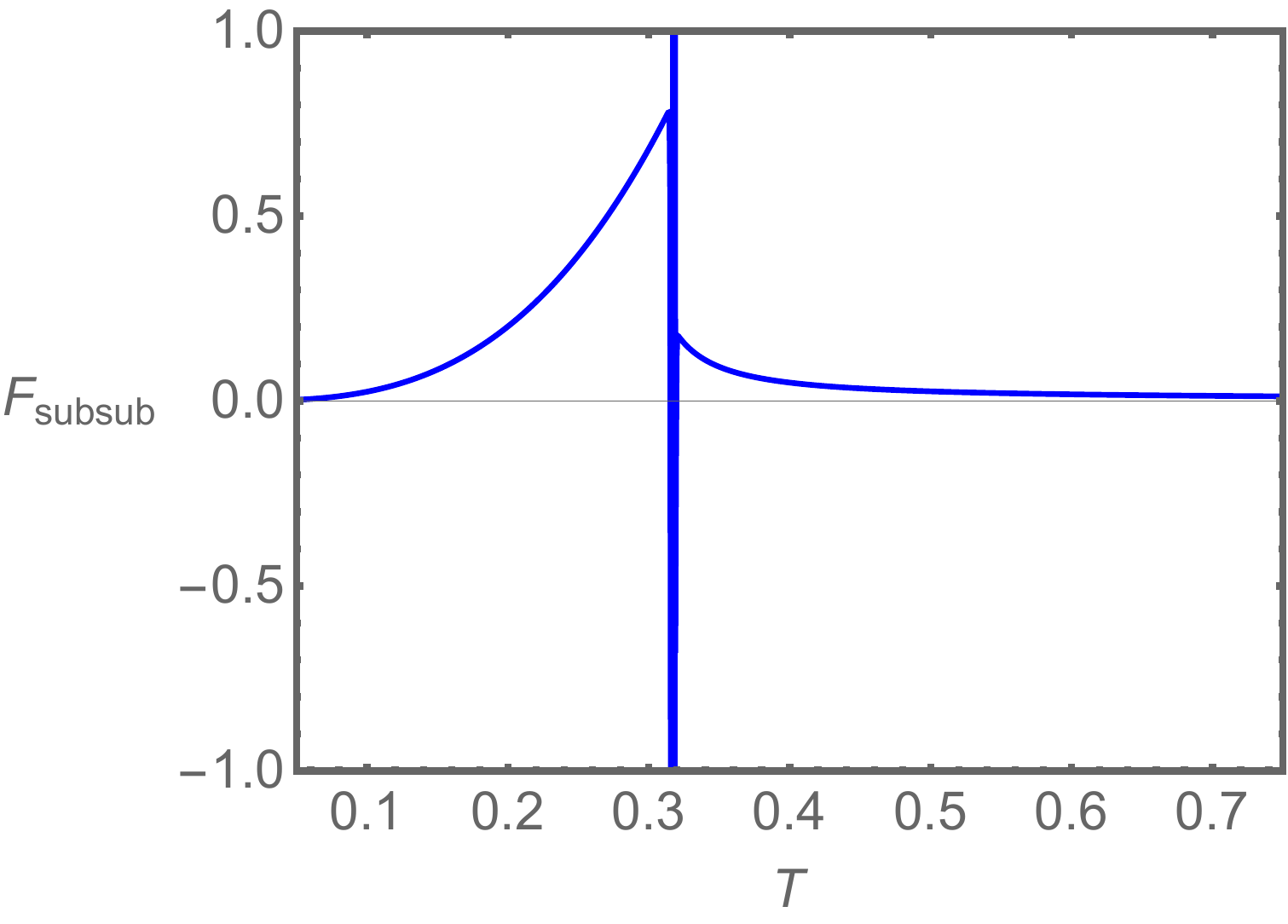}
  \caption{An illustration of the subsubleading-order free energy of the Schwarzschild-AdS spacetime.}\label{subsub}%\vspace{-2em}
\end{figure}

From the above calculation, the subleading-order contribution of $G_\tN$ can be derived by expanding around the classical saddle, which can be verified by numerical demonstration of the leading and subleading-order contributions.
Figure \ref{sub} illustrates the numerical result of the subleading free energies of the Schwarzschild-AdS and RN-AdS spacetimes. As can be seen from the figures, away from the choppy transition points, the subleading contributions of the black holes are exactly $F_{\text{sub}}=T/2$ for both the Schwarzschild-AdS and RN-AdS black holes. There must be some universal structures for the two cases that result in the same subleading-order behavior. 
Moreover, the thermal AdS phase exactly matches the analytical analysis, which is $F_{\text{sub}}=T$.

The subsubleading-order correction to the black hole free energy can be derived and shown in Fig. \ref{subsub}. Besides the vibration near the transition point, we have different smooth behaviors before and after the transition, which can be fitted as a function of ensemble temperature $T$. The physical free energy should be the summation of those different pieces of the free energies, i.e.
\be
\ave{F}=F_{\text{leading}}+F_{\text{sub}}\cdot G_\tN^0+F_{\text{subsub}}\cdot G_\tN^1+\mathcal{O}(G_\tN^2)\,,\label{expand}
\ee
with the black hole free energy $F_{\text{leading}}\propto 1/G_\tN$. 
For small $G_\tN$, such as $G_\tN=1/1000$, $\mathcal{O}(G_\tN^2)$ terms can be ignored, and keeping the first several terms in \eqref{expand} is good enough to capture the corrected thermodynamics due to the geometries with conical singularity.

\section{Summary and discussion}
\label{con}

In this letter, the action for Euclidean geometries with a conical singularity was calculated, and we argued that in the path integral approach to quantum gravity, all those geometries should be included in the path integral.
Because the geometries are described by the same Lorentzian metric, any physical quantity related to the metric should be ensemble-averaged over the Euclidean geometries.
Working with averaged quantities, we concluded that the Hawking-Page transition for the Schwarzschild-AdS spacetime and the small-large black hole transition for the RN-AdS spacetime can be directly derived in the semi-classical limit. No extra assumption is needed.
The averaged free energy deviates from that of the black hole for relatively large $G_\tN$. 
The subleading order of $G_\tN$ contributions can be reproduced by expanding around the saddle point, which are both $T/2$ for the Schwarzschild and RN black holes. The analytical and numerical results exactly match.
Thus, we conjecture that there is a universal structure for black holes and the same result should also hold for other black holes. 
Beyond the semi-classical limit, we have quantum-corrected black hole thermodynamics, where the subleading contributions can not be ignored and the effects considered here would be important.

We have some further discussions related to the main results. In the Schwarzschild-AdS case, we did not mention the thermal AdS but just evaluated the averaged free energy of the Schwarzschild-AdS metric. It turns out that pure AdS naturally arises and dominates at low temperatures, as can be seen from Fig. \ref{SAdS}. However, for Einstein-Maxwell theory, the thermal AdS is not a solution for non-vanishing $Q$, we would not see such a phase transition between black holes and pure AdS. The Hawking-Page and small-large black hole phase transitions are two types of transitions, but the ensemble average seems clever enough to understand the two types simultaneously.

Note that the subleading correction of the free energy from ensemble-averaged theory should be universal, which does not depend on the spacetime geometry and dimensionality. 
The claim was checked for the Kerr-AdS black hole \cite{Cheng:2024efw} and the Einstein-Gauss-Bonnet black holes in five and six dimensions \cite{Ali:2024adt}. For those spacetimes, the subleading-order terms are all half of the temperature. 
The physical reason for the universality is that we are using a Gaussian distribution of Euclidean geometries when we deviate away from the classical saddle points. However for larger $G_\tN$, when more subsubleading terms are needed, Gaussian distribution may not be an ideal approximation anymore, as can be seen from Fig. \ref{Pdistribution}. In such a case, we would see the signature of each geometry from those subsubleading-order terms.

For the Hawking-Page phase transition, there is a dual large-$N$ SU($N$) description. It is well known that it is a transition between states with free energies scaling with $\mathcal{O}(N^0)$ and $\mathcal{O}(N^2)$ in the boundary theory. 
The thermal AdS spacetime is dual to state with constant free energy and the asymptotic AdS black hole is dual to state with free energy scales as $N^2$.
The finite $G_\tN$ effects that deviate from the thermal AdS and black hole, should dual to the boundary finite $N$ physics. 
One can figure out the subleading terms in large $N$ expansion through the duality between the two sides, which may be able to provide further insights into the boundary theory.
It was shown that the Hawking-Page transition can be reproduced on a spin chain with an average over the randomized coupling \cite{Perez-Garcia:2024pcq}. Due to the similarity, it would be interesting to find possible connections.

\begin{acknowledgments}
\noindent \textbf{Acknowledgments}\\
We would like to thank Hong L\"{u}, Yang An, Bum-Hoon Lee, Pujian Mao, Robert Mann, Si-Jiang Yang, and Jin Wang for their helpful discussions. 
This work is supported by the National Natural Science Foundation of China (NSFC) under Grant No. 12405073, No. 12475056, No. 12247101, and No. 12475055, and the 111 Project under Grant No. B20063.
\end{acknowledgments}

%%%%%%%%%%%%%%%%%%%%%%%%%%%%%%%%%%%%%%%%%%%%%%%%%%%%%%%%%%%%%%%%%%%%%%%%%%%%%%%%%%%%%%%%%%%%%%%s%%%%%

\providecommand{\href}[2]{#2}\begingroup\raggedright\endgroup

%%%%%%%%%%%%%%%%%%%%%%%%%%%%%%%%%%%%%%%%%%%%%%%%%%%%%%%%%%%%%%%%%%%%%%%%%%%%%%%%%%%%%%%%%%%%%%%%%%%%

%\newpage
\appendix
\section{Density matrix and probability distribution}
\label{appA}

\begin{figure*}[hbt]
    \centering
\subfloat[Probability distribution with $G_\tN=1/10$]{\includegraphics[width=0.45\linewidth]{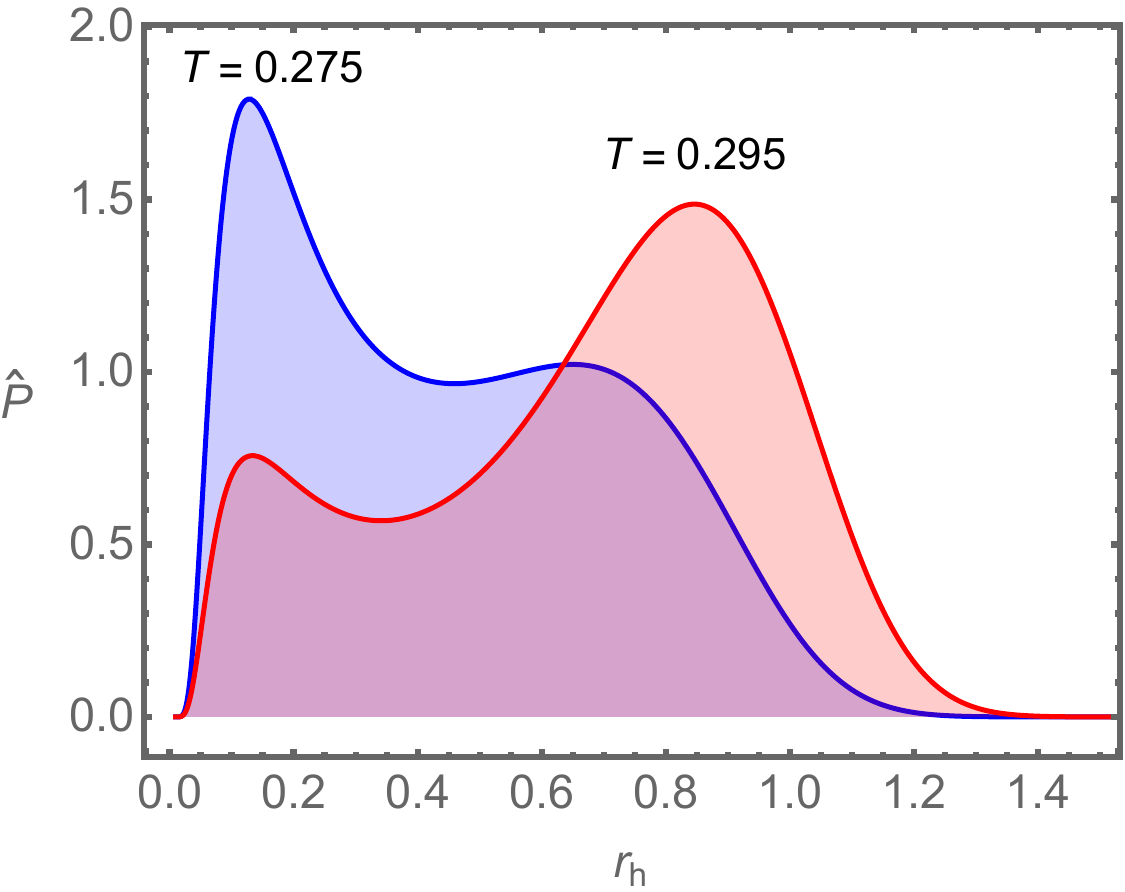}\label{Pdistribution-a}}~~~~
\subfloat[Probability distribution with $G_\tN=1/1000$]{\includegraphics[width=0.45\linewidth]{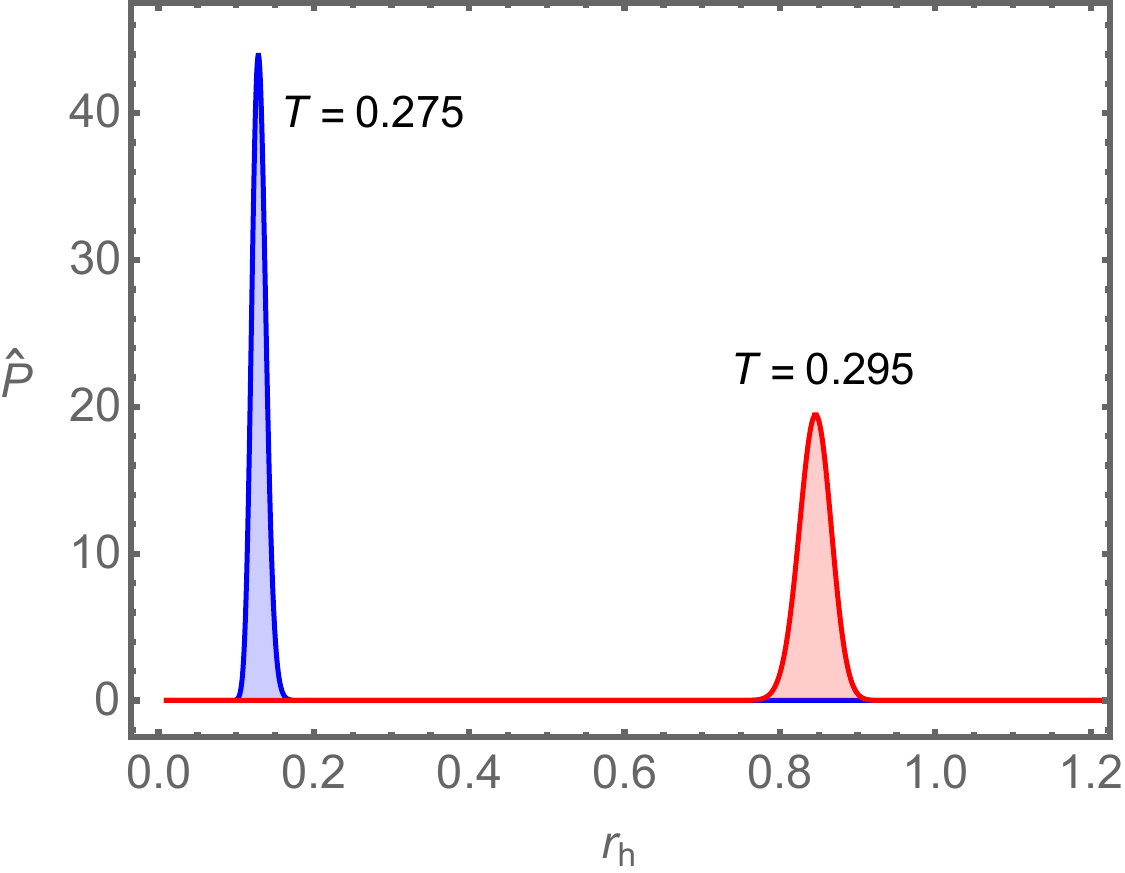}\label{Pdistribution-b}}
    \caption{Probability distributions for the RN-AdS case with different values of $G_\tN$. The blue and red curves represent the small black hole phase with $T = 0.275$ and the large black hole phase with $T = 0.295$, respectively.}
    \label{Pdistribution}
\end{figure*}

One of the customary approaches to a statistical interpretation of thermodynamics is the method of Euclidean path integral. 
We review the method and apply it to black hole thermodynamics here.

For scalar fields, similar to the path integral formula of the transition amplitude, the density matrix and partition function also admit a Euclidean path integral description:
\bea
\bra{\phi_b}e^{-\beta H}\ket{\phi_a}=\int_{\phi(\textbf{x},0)=\phi_a(\textbf{x})}^{\phi(\textbf{x},\beta)=\phi_b(\textbf{x})}[\mathcal{D} \phi]~ e^{-I_{\text{E}}}\,,\label{density}
\eea
%and
\bea
Z_{\phi}=\sum_{\phi}\bra{\phi}e^{-\beta H}\ket{\phi}=\int_{\text{periodic}}[\mathcal{D} \phi]~ e^{-I_{\text{E}}}\,,\label{Zphi}
\eea
where $I_\tE$ is the Euclidean action and the periodic boundary condition is along the $\tau$ direction in the partition function calculation.
The path integral representation implies the weight of each state can be evaluated through the path integral, as far as the Euclidean action can be calculated.

The method can be generalized to gravitational physics \cite{Gibbons:1976ue}.
Moreover, the density matrix for states, including the ones with conical singularity on their Euclidean counterpart, can be calculated through Euclidean path integral
\be
\rho=\sum_{
\begin{matrix}
\begin{tikzpicture}
\draw[thick] (-0.05,0) .. controls (-0.05,0) and (-0.175,0.0125) .. (-0.3,0.125);
\draw[thick] (-0.05,0.25) .. controls (-0.05,0.25) and (-0.175,1.9/8) .. (-0.3,0.125);
\draw[thick] (-0.05,0.125) ellipse (0.125/2 and 0.25/2);
\fill[red] (-0.3,0.125) circle (1pt);
\end{tikzpicture}
\end{matrix}
}
e^{-I_\tE[
\begin{matrix}
\begin{tikzpicture}
\draw[thick] (0.5/2,0) .. controls (0.5/2,0) and (0.25/2,0.025/2) .. (0,0.25/2);
\draw[thick] (0.5/2,0.5/2) .. controls (0.5/2,0.5/2) and (0.25/2,1.9/8) .. (0,0.25/2);
\draw[thick] (0.5/2,0.25/2) ellipse (0.125/2 and 0.25/2);
\fill[red] (0,0.25/2) circle (1pt);
\end{tikzpicture}
\end{matrix}
]}
\ket{
\begin{matrix}
\begin{tikzpicture}
\draw[thick] (0.5,0) .. controls (0.5,0) and (0.25,0.025) .. (0,0.25);
\draw[thick] (0.5,0.5) .. controls (0.5,0.5) and (0.25,1.9/4) .. (0,0.25);
\draw[thick] (0.5,0.25) ellipse (0.125 and 0.25);
\fill[red] (0,0.25) circle (1pt);
\end{tikzpicture}
\end{matrix}
}
\bra{
\begin{matrix}
\begin{tikzpicture}
\draw[thick] (-0.5,0) .. controls (-0.5,0) and (-0.25,-0.025) .. (0,-0.25);
\draw[thick] (-0.5,-0.5) .. controls (-0.5,-0.5) and (-0.25,-1.9/4) .. (0,-0.25);
\draw[thick] (-0.5,-0.25) ellipse (-0.125 and -0.25);
\fill[red] (0,-0.25) circle (1pt);
\end{tikzpicture}	
\end{matrix}
}.
\ee
With the density matrix, the free energy associated with any metric should be defined as an ensemble-averaged one
\be
\ave{F}=\frac{1}{Z}\tr(\rho F_{\text{gen}})\,,
\ee
with the partition function $Z=\tr~\rho$ and the generalized free energy $F_\text{gen}=I_\tE/\beta$. The normalized weight of each state $\hat{P}$ in the phase space, represented by the diagonal component of the density matrix divided by $Z$, can be calculated by
\be
\hat{P}[\begin{matrix}
\begin{tikzpicture}
\draw[thick] (0.5/2,0) .. controls (0.5/2,0) and (0.25/2,0.025/2) .. (0,0.25/2);
\draw[thick] (0.5/2,0.5/2) .. controls (0.5/2,0.5/2) and (0.25/2,1.9/8) .. (0,0.25/2);
\draw[thick] (0.5/2,0.25/2) ellipse (0.125/2 and 0.25/2);
\fill[red] (0,0.25/2) circle (1pt);
\end{tikzpicture}
\end{matrix}]
%=\frac{P[\begin{matrix}
%\begin{tikzpicture}
%\draw[thick] (0.5/2,0) .. controls (0.5/2,0) and (0.25/2,0.025/2) .. (0,0.25/2);
%\draw[thick] (0.5/2,0.5/2) .. controls (0.5/2,0.5/2) and (0.25/2,1.9/8) .. (0,0.25/2);
%\draw[thick] (0.5/2,0.25/2) ellipse (0.125/2 and 0.25/2);
%\fill[red] (0,0.25/2) circle (1pt);
%\end{tikzpicture}
%\end{matrix}]}{Z}
=\frac{e^{-I_\tE[\begin{matrix}
\begin{tikzpicture}
\draw[thick] (0.5/2,0) .. controls (0.5/2,0) and (0.25/2,0.025/2) .. (0,0.25/2);
\draw[thick] (0.5/2,0.5/2) .. controls (0.5/2,0.5/2) and (0.25/2,1.9/8) .. (0,0.25/2);
\draw[thick] (0.5/2,0.25/2) ellipse (0.125/2 and 0.25/2);
\fill[red] (0,0.25/2) circle (1pt);
\end{tikzpicture}
\end{matrix}]}}{Z}\,.
\ee

In the case being considered here, the weight of each state is labeled by a continuous parameter $r_\hh$, so we have a probability distribution of states. 
Taking the RN-AdS case as an example, we plot the probability distributions of the RN-AdS case with small and large $G_\tN$ in Fig. \ref{Pdistribution}.
There are three different situations:
\begin{itemize}
    \item For relatively large $G_\tN$ as shown in Fig. \ref{Pdistribution-a}, the geometries away from the global maximum have significant contribution. 
    We can numerically integrate over all the geometries with the probability distribution given in Fig. \ref{Pdistribution-a}. Correspondingly, the ensemble-averaged result reproduces the finite $G_\tN$ curves shown in Fig. \ref{ave}.
    \item For small but finite $G_\tN$ as shown in Fig. \ref{Pdistribution-b}, the whole distribution can be approximated by a Gaussian distribution. With the approximation, the ensemble average can be carried out analytically, as described in section IV on non-classical corrections to black hole thermodynamics.
     It turns out that the analytical result contains the black hole result proportional to $1/G_\tN$ as the leading-order contribution and the $G_\tN^0$ contribution as the subleading-order one.
    \item For the case $G_\tN \to 0$, the probability distribution turns to a Dirac delta function, which picks up one single geometry in the Euclidean path integral. This explains the black hole phase transition.
\end{itemize}

It is worth emphasizing that for small $G_\tN$ shown in Fig. \ref{Pdistribution-b}, the Gaussian distribution can be obtained by expanding the original distribution to the second-order of the distance $\Delta$ near the Hawking saddle.
In principle, one should integrate over the whole domain of $r_\hh$. Here, we believe that integrating over $(-5\sigma,5\sigma)$ only contains a negligible error and is good enough.
So, the analytical calculation in section IV should not be regarded as a perturbation result. The proper way of interpreting it is that we can use a Gaussian distribution to approximate the whole distribution for small $G_\tN$, and the integration gives out the correct subleading contribution.


\begin{thebibliography}{100}

\bibitem{Hawking:1982dh}
S.~W. Hawking and D.~N. Page, ``{Thermodynamics of Black Holes in anti-De
  Sitter Space},'' \href{http://dx.doi.org/10.1007/BF01208266}{{\em Commun.
  Math. Phys.} {\bfseries 87} (1983) 577--588}.

\bibitem{Witten:1998zw}
E.~Witten, ``{Anti-de Sitter space, thermal phase transition, and confinement
  in gauge theories},''
  \href{http://dx.doi.org/10.4310/ATMP.1998.v2.n3.a3}{{\em Adv. Theor. Math.
  Phys.} {\bfseries 2} (1998) 505--532},
  \href{http://arxiv.org/abs/hep-th/9803131}{{\ttfamily arXiv:hep-th/9803131}}.
  
\bibitem{Gross:1980he}
D.~J.~Gross and E.~Witten,
``Possible Third Order Phase Transition in the Large N Lattice Gauge Theory,''
\href{http://dx.doi.org/10.1103/PhysRevD.21.446}{{\em Phys. Rev. D} {\bfseries 21} (1980) 446--453}.
%925 citations counted in INSPIRE as of 05 May 2024

%\cite{Sundborg:1999ue}
\bibitem{Sundborg:1999ue}
B.~Sundborg,
``The Hagedorn transition, deconfinement and N=4 SYM theory,''
\href{http://dx.doi.org/10.1016/S0550-3213(00)00044-4}{{\em Nucl. Phys. B}
  {\bfseries 573} (2000) 349--363},
  \href{http://arxiv.org/abs/hep-th/9908001}{{\ttfamily arXiv:hep-th/9908001 [hep-th]}}.
%413 citations counted in INSPIRE as of 05 May 2024

\bibitem{Chamblin:1999hg}
A.~Chamblin, R.~Emparan, C.~V. Johnson, and R.~C. Myers, ``{Holography,
  thermodynamics and fluctuations of charged AdS black holes},''
  \href{http://dx.doi.org/10.1103/PhysRevD.60.104026}{{\em Phys. Rev. D}
  {\bfseries 60} (1999) 104026},
  \href{http://arxiv.org/abs/hep-th/9904197}{{\ttfamily arXiv:hep-th/9904197}}.

\bibitem{Chamblin:1999tk}
A.~Chamblin, R.~Emparan, C.~V. Johnson, and R.~C. Myers, ``{Charged AdS black
  holes and catastrophic holography},''
  \href{http://dx.doi.org/10.1103/PhysRevD.60.064018}{{\em Phys. Rev. D}
  {\bfseries 60} (1999) 064018},
  \href{http://arxiv.org/abs/hep-th/9902170}{{\ttfamily arXiv:hep-th/9902170}}.

\bibitem{Kastor:2009wy}
D.~Kastor, S.~Ray, and J.~Traschen, ``{Enthalpy and the Mechanics of AdS Black
  Holes},'' \href{http://dx.doi.org/10.1088/0264-9381/26/19/195011}{{\em Class.
  Quant. Grav.} {\bfseries 26} (2009) 195011},
  \href{http://arxiv.org/abs/0904.2765}{{\ttfamily arXiv:0904.2765 [hep-th]}}.

\bibitem{Dolan:2011xt}
B.~P. Dolan, ``{Pressure and volume in the first law of black hole
  thermodynamics},''
  \href{http://dx.doi.org/10.1088/0264-9381/28/23/235017}{{\em Class. Quant.
  Grav.} {\bfseries 28} (2011) 235017},
  \href{http://arxiv.org/abs/1106.6260}{{\ttfamily arXiv:1106.6260 [gr-qc]}}.

\bibitem{Kubiznak:2012wp}
D.~Kubiznak and R.~B. Mann, ``{P-V criticality of charged AdS black holes},''
  \href{http://dx.doi.org/10.1007/JHEP07(2012)033}{{\em JHEP} {\bfseries 07}
  (2012) 033}, \href{http://arxiv.org/abs/1205.0559}{{\ttfamily arXiv:1205.0559
  [hep-th]}}.

\bibitem{Wei:2015iwa}
S.-W. Wei and Y.-X. Liu, ``{Insight into the Microscopic Structure of an AdS
  Black Hole from a Thermodynamical Phase Transition},''
  \href{http://dx.doi.org/10.1103/PhysRevLett.115.111302}{{\em Phys. Rev.
  Lett.} {\bfseries 115} (2015) 111302},
  \href{http://arxiv.org/abs/1502.00386}{{\ttfamily arXiv:1502.00386 [gr-qc]}}.
  [Erratum: Phys.Rev.Lett. {\bfseries 116} (2016) 169903].

\bibitem{Li:2020khm}
R.~Li and J.~Wang, ``{Thermodynamics and kinetics of Hawking-Page phase
  transition},'' \href{http://dx.doi.org/10.1103/PhysRevD.102.024085}{{\em
  Phys. Rev. D} {\bfseries 102} (2020) 024085}.

\bibitem{Li:2020nsy}
R.~Li, K.~Zhang, and J.~Wang, ``{Thermal dynamic phase transition of
  Reissner-Nordstr\"om Anti-de Sitter black holes on free energy landscape},''
  \href{http://dx.doi.org/10.1007/JHEP10(2020)090}{{\em JHEP} {\bfseries 10}
  (2020) 090}, \href{http://arxiv.org/abs/2008.00495}{{\ttfamily
  arXiv:2008.00495 [hep-th]}}.

\bibitem{Li:2020spm}
R.~Li and J.~Wang, ``{Energy and entropy compensation, phase transition and
  kinetics of four-dimensional charged Gauss-Bonnet Anti-de Sitter black holes
  on the underlying free energy landscape},''
  \href{http://dx.doi.org/10.1016/j.nuclphysb.2022.115714}{{\em Nucl. Phys. B}
  {\bfseries 976} (2022) 115714},
  \href{http://arxiv.org/abs/2012.05424}{{\ttfamily arXiv:2012.05424 [gr-qc]}}.

\bibitem{Yang:2021ljn}
S.-J. Yang, R.~Zhou, S.-W. Wei, and Y.-X. Liu, ``{Kinetics of a phase
  transition for a Kerr-AdS black hole on the free-energy landscape},''
  \href{http://dx.doi.org/10.1103/PhysRevD.105.084030}{{\em Phys. Rev. D}
  {\bfseries 105} (2022) 084030},
  \href{http://arxiv.org/abs/2105.00491}{{\ttfamily arXiv:2105.00491 [gr-qc]}}.

\bibitem{Li:2021vdp}
R.~Li, K.~Zhang and J.~Wang,
``Probing black hole microstructure with the kinetic turnover of phase transition,''
\href{http://dx.doi.org/10.1103/PhysRevD.104.084076}{{\em Phys. Rev. D}
  {\bfseries 104} (2021) 084076},
  \href{http://arxiv.org/abs/2102.09439}{{\ttfamily arXiv:2102.09439 [gr-qc]}}.
%22 citations counted in INSPIRE as of 08 Aug 2024

%\cite{Li:2022yti}
\bibitem{Li:2022yti}
R.~Li and J.~Wang,
``Non-Markovian dynamics of black hole phase transition,''
\href{http://dx.doi.org/10.1103/PhysRevD.106.104039}{{\em Phys. Rev. D}
  {\bfseries 106} (2022) 104039},
  \href{http://arxiv.org/abs/2205.00594}{{\ttfamily arXiv:2205.00594 [gr-qc]}}.
%8 citations counted in INSPIRE as of 08 Aug 2024


%\cite{Li:2022ylz}
\bibitem{Li:2022ylz}
R.~Li and J.~Wang,
``Kinetics of Hawking-Page phase transition with the non-Markovian effects,''
\href{http://dx.doi.org/10.1007/JHEP05(2022)128}{{\em JHEP}
  {\bfseries 05} (2022) 128},
  \href{http://arxiv.org/abs/2201.06138}{{\ttfamily arXiv:2201.06138 [gr-qc]}}.
%7 citations counted in INSPIRE as of 08 Aug 2024

\bibitem{Li:2022oup}
R.~Li and J.~Wang, \emph{{Generalized free energy landscape of a black hole
  phase transition}},
  \href{http://dx.doi.org/10.1103/PhysRevD.106.106015}{\emph{Phys. Rev. D} {\bf
  106} (2022) 106015}, \href{http://arxiv.org/abs/2206.02623}{{\tt
  arXiv:2206.02623 [hep-th]}}.

\bibitem{Gibbons:1976ue}
G.~W. Gibbons and S.~W. Hawking, ``{Action Integrals and Partition Functions in
  Quantum Gravity},'' \href{http://dx.doi.org/10.1103/PhysRevD.15.2752}{{\em
  Phys. Rev. D} {\bfseries 15} (1977) 2752--2756}.

\bibitem{York:1986it}
J.~W. York, Jr., \emph{{Black hole thermodynamics and the Euclidean Einstein
  action}}, \href{http://dx.doi.org/10.1103/PhysRevD.33.2092}{\emph{Phys. Rev.
  D} {\bf 33} (1986) 2092--2099}.

\bibitem{Braden:1987ad}
H.~W. Braden, B.~F. Whiting and J.~W. York, Jr., \emph{{Density of States for
  the Gravitational Field in Black Hole Topologies}},
  \href{http://dx.doi.org/10.1103/PhysRevD.36.3614}{\emph{Phys. Rev. D} {\bf
  36} (1987) 3614}.

\bibitem{Whiting:1988qr}
B.~F. Whiting and J.~W. York, Jr., \emph{{Action Principle and Partition
  Function for the Gravitational Field in Black Hole Topologies}},
  \href{http://dx.doi.org/10.1103/PhysRevLett.61.1336}{\emph{Phys. Rev. Lett.}
  {\bf 61} (1988) 1336}.

\bibitem{York:1988va}
J.~W. York, Jr., \emph{{Action and Free Energy for Black Hole Topologies}},
  \href{http://dx.doi.org/10.1016/0378-4371(89)90540-2}{\emph{Physica A} {\bf
  158} (1989) 425--436}.

%\cite{Solodukhin:1994yz}
\bibitem{Solodukhin:1994yz}
S.~N.~Solodukhin,
``The Conical singularity and quantum corrections to entropy of black hole,''
\href{http://dx.doi.org/10.1103/PhysRevD.51.609}{{\em Phys. Rev. D}
  {\bfseries 51} (1995) 609--617}, \href{http://arxiv.org/abs/hep-th/9407001}{{\ttfamily arXiv:hep-th/9407001}}.
%274 citations counted in INSPIRE as of 22 Jul 2024

\bibitem{Fursaev:1995ef}
D.~V. Fursaev and S.~N. Solodukhin, ``{On the description of the Riemannian
  geometry in the presence of conical defects},''
  \href{http://dx.doi.org/10.1103/PhysRevD.52.2133}{{\em Phys. Rev. D}
  {\bfseries 52} (1995) 2133--2143},
  \href{http://arxiv.org/abs/hep-th/9501127}{{\ttfamily arXiv:hep-th/9501127}}.

\bibitem{Mann:1996bi}
R.~B. Mann and S.~N. Solodukhin, ``{Conical geometry and quantum entropy of a
  charged Kerr black hole},''
  \href{http://dx.doi.org/10.1103/PhysRevD.54.3932}{{\em Phys. Rev. D}
  {\bfseries 54} (1996) 3932--3940}, \href{http://arxiv.org/abs/hep-th/9604118}{{\ttfamily arXiv:hep-th/9604118}}.

\bibitem{Solodukhin:2011gn}
S.~N. Solodukhin, \emph{{Entanglement entropy of black holes}},
  \href{http://dx.doi.org/10.12942/lrr-2011-8}{\emph{Living Rev. Rel.} {\bf 14}
  (2011) 8}, \href{http://arxiv.org/abs/1104.3712}{{\tt arXiv:1104.3712 [hep-th]}}.

\bibitem{Carlip:1993sa}
S.~Carlip and C.~Teitelboim, \emph{{The Off-shell black hole}},
  \href{http://dx.doi.org/10.1088/0264-9381/12/7/011}{\emph{Class. Quant.
  Grav.} {\bf 12} (1995) 1699--1704},
  \href{http://arxiv.org/abs/gr-qc/9312002}{{\tt arXiv:gr-qc/9312002}}.



\bibitem{Saad2019}
P.~Saad, S.~H. Shenker, and D.~Stanford, ``JT gravity as a matrix integral,''
  \href{http://arxiv.org/abs/1903.11115}{{\ttfamily arXiv:1903.11115
  [hep-th]}}.

\bibitem{AfkhamiJeddi2020}
N.~Afkhami-Jeddi, H.~Cohn, T.~Hartman, and A.~Tajdini, ``Free partition
  functions and an averaged holographic duality,''
  \href{http://dx.doi.org/10.1007/JHEP01(2021)130}{\textit{JHEP} \textbf{01} (2021) 130}, \href{http://arxiv.org/abs/2006.04839}{{\ttfamily
  arXiv:2006.04839 [hep-th]}}.

\bibitem{Maloney2020}
A.~Maloney and E.~Witten, ``Averaging Over Narain Moduli Space,''
\href{http://dx.doi.org/10.1007/JHEP10(2020)187}{\textit{JHEP} \textbf{10} (2020) 187},
  \href{http://arxiv.org/abs/2006.04855}{{\ttfamily arXiv:2006.04855
  [hep-th]}}.


\bibitem{Penington2019a}
G.~Penington, S.~H. Shenker, D.~Stanford, and Z.~Yang, ``Replica wormholes and
  the black hole interior,''
  \href{http://dx.doi.org/10.1007/JHEP03(2022)205}{\textit{JHEP} \textbf{03} (2022) 205},
  \href{http://arxiv.org/abs/1911.11977v2}{{\ttfamily arXiv:1911.11977v2}}.
  
\bibitem{Marolf2020a}
D.~Marolf and H.~Maxfield, ``Transcending the ensemble: baby universes,
  spacetime wormholes, and the order and disorder of black hole information,''
  \href{http://dx.doi.org/10.1007/jhep08(2020)044}{{\em JHEP}
  {\bfseries 08} (2020) 044},
  \href{http://arxiv.org/abs/2002.08950}{{\ttfamily arXiv:2002.08950}}.

\bibitem{Maldacena2016a}
J.~Maldacena and D.~Stanford, ``Comments on the Sachdev-Ye-Kitaev model,''
  \href{http://dx.doi.org/10.1103/PhysRevD.94.106002}{{\em Phys. Rev. D }\textbf{94} (2016) 106002 },
  \href{http://arxiv.org/abs/1604.07818}{{\ttfamily arXiv:1604.07818
  [hep-th]}}.

\bibitem{Jensen2016}
K.~Jensen, ``Chaos in AdS$_2$ holography,''
  \href{http://dx.doi.org/10.1103/PhysRevLett.117.111601}{{\em Phys. Rev. Lett. }\textbf{117} (2016) 111601},
  \href{http://arxiv.org/abs/1605.06098}{{\ttfamily arXiv:1605.06098
  [hep-th]}}.

\bibitem{Heckman:2021vzx}
J.~J. Heckman, A.~P. Turner, and X.~Yu, ``{Disorder averaging and its UV
  discontents},'' \href{http://dx.doi.org/10.1103/PhysRevD.105.086021}{{\em
  Phys. Rev. D} {\bfseries 105} (2022) 086021},
  \href{http://arxiv.org/abs/2111.06404}{{\ttfamily arXiv:2111.06404
  [hep-th]}}.

%\cite{Cheng:2022nra}
\bibitem{Cheng:2022nra}
P.~Cheng and P.~Mao,
``Notes on Wormhole Cancellation and Factorization,''
\href{http://doi.org/10.1140/epjc/s10052-024-13045-5}{{\em
  Eur. Phys. J. C} {\bfseries 84} (2024) 675},
 \href{http://arxiv.org/abs/2208.08456}{{\ttfamily arXiv:2208.08456
  [hep-th]}}.
%3 citations counted in INSPIRE as of 14 May 2024

%\cite{Jafferis:2024jkb}
\bibitem{Jafferis:2024jkb}
D.~L.~Jafferis, L.~Rozenberg and G.~Wong,
``3d Gravity as a random ensemble,''
 \href{http://arxiv.org/abs/2407.02649}{{\ttfamily arXiv:2407.02649
  [hep-th]}}.
%1 citations counted in INSPIRE as of 21 Jul 2024

\bibitem{Bardeen:1973gs}
J.~M. Bardeen, B.~Carter, and S.~W. Hawking, ``{The four laws of black hole
  mechanics},'' \href{http://dx.doi.org/10.1007/BF01645742}{{\em Commun. Math.
  Phys.} {\bfseries 31} (1973) 161--170}.

\bibitem{Bekenstein:1973ur}
J.~D. Bekenstein, ``{Black holes and entropy},''
  \href{http://dx.doi.org/10.1103/PhysRevD.7.2333}{{\em Phys. Rev. D}
  {\bfseries 7} (1973) 2333--2346}.

\bibitem{Hawking:1975vcx}
S.~W. Hawking, ``{Particle Creation by Black Holes},''
  \href{http://dx.doi.org/10.1007/BF02345020}{{\em Commun. Math. Phys.}
  {\bfseries 43} (1975) 199--220}. [Erratum: Commun.Math.Phys. {\bfseries 46} (1976) 206].

\bibitem{Hawking:1976de}
S.~W. Hawking, ``{Black Holes and Thermodynamics},''
  \href{http://dx.doi.org/10.1103/PhysRevD.13.191}{{\em Phys. Rev. D}
  {\bfseries 13} (1976) 191--197}.

%\cite{Cheng:2024efw}
\bibitem{Cheng:2024efw}
P.~Cheng, J.~Pan, H.~Xu and S.~J.~Yang,
``Thermodynamics of the Kerr-AdS black hole from an ensemble-averaged theory,''
\href{http://arxiv.org/abs/2410.23006}{{\ttfamily arXiv:2410.23006 [hep-th]}}.
%1 citations counted in INSPIRE as of 15 Nov 2024

%\cite{Ali:2024adt}
\bibitem{Ali:2024adt}
M.~S.~Ali, C.~Fairoos, C.~L.~A.~Rizwan, T.~K.~Safir and P.~Cheng,
%``Thermodynamics of Einstein-Gauss-Bonnet Black Holes and Ensemble-averaged Theory,''
\href{http://arxiv.org/abs/2411.07147}{{\ttfamily arXiv:2411.07147 [gr-qc]}}.
%0 citations counted in INSPIRE as of 15 Nov 2024
  
  %\cite{Perez-Garcia:2024pcq}
\bibitem{Perez-Garcia:2024pcq}
D.~P\'erez-Garc\'\i{}a, L.~Santilli and M.~Tierz,
``Hawking-Page transition on a spin chain,''
\href{http://dx.doi.org/10.1103/PhysRevResearch.6.033007}{{\em Phys. Rev. Res.}
  {\bfseries 6} (2024) 033007},
  \href{http://arxiv.org/abs/2401.13963}{{\ttfamily arXiv:2401.13963 [quant-ph]}}.
%1 citations counted in INSPIRE as of 26 Aug 2024

\end{thebibliography}
\end{document}